# Diversity of Expertise is Key to Scientific Impact: a Large-Scale Analysis in the Field of Computer Science

Angelo Salatino[*], Simone Angioni[**], Francesco Osborne[***], Diego Reforgiato Recupero[**], and Enrico Motta[*]

[*]*angelo.salatino@open.ac.uk; enrico.motta@open.ac.uk*
0000-0002-4763-3943; 0000-0003-0015-1952
Knowledge Media Institute, The Open University, United Kingdom

[**]*simone.angioni@unica.it; diego.reforgiato@unica.it*
0000-0002-6682-3419, 0000-0001-8646-6183
Department of Mathematics and Computer Science, University of Cagliari, Italy

[***]*francesco.osborne@open.ac.uk*
0000-0001-6557-3131
Knowledge Media Institute, The Open University, United Kingdom;
Department of Business and Law, University of Milano Bicocca, Milan, Italy

**Abstract:** Understanding the relationship between the composition of a research team and the potential impact of their research papers is crucial as it can steer the development of new science policies for improving the research enterprise. Numerous studies assess how the characteristics and diversity of research teams can influence their performance across several dimensions: ethnicity, internationality, size, and others. In this paper, we explore the impact of diversity in terms of the authors' expertise. To this purpose, we retrieved 114K papers in the field of Computer Science and analysed how the diversity of research fields within a research team relates to the number of citations their papers received in the upcoming 5 years. The results show that two different metrics we defined, reflecting the diversity of expertise, are significantly associated with the number of citations. This suggests that, at least in Computer Science, diversity of expertise is key to scientific impact.

## 1. Introduction

Understanding the relationship between the composition of a research team and the potential impact of their research papers is crucial as it can lead to the development of science policies and best practices to drive innovation forward. A commonly examined characteristic is the diversity of the team across a number of dimensions, such as ethnicity (AlShebli et al., 2018), gender (Nielsen et al., 2017), disciplinary backgrounds (Uzzi et al., 2013), team size (Wu et al., 2019), and others. Less attention had been paid to the expertise diversity of the researchers. However, over the past few years, there has been a growing emphasis among funding agencies, scientific journals, and government institutions on the importance of interdisciplinary approaches and collaboration between scientific fields. While the present landscape may lead researchers to become overly specialized in narrow fields of study, the scientific community aspires to unite in their efforts to address extensive societal challenges, such as climate change, poverty, disease, inequality, and the imperative for sustainable development. By their very nature, these challenges demand complex and multifaced solutions that necessitate the integration of diverse expertise. The cross-pollination of ideas from different expertise may also break down traditional silos between disciplines and uncover unexpected insights that can drive new discoveries.

In this paper, we present a scientometric analysis in which we assess whether a diverse pool of expertise within a research team can influence their scientific impact, measured as the number of citations received by the resulting research papers in the upcoming 5 years.

The analysis was performed on 114,203 Computer Science papers from the Academia/Industry DynAmics (AIDA) Knowledge Graph [1], published within the 2010-2015 timeframe. To assess the diversity of a team, we characterise a researcher's expertise as the distribution of research topics of their paper in the previous 5 years. To this purpose, we leverage the Computer Science Ontology, which consists of 14K topics and provides a more fine-grained representation compared to the generic disciplines provided by typical scholarly datasets such as Scopus and Web of Science. We then computed the pairwise cosine similarity between each couple of authors in a paper and defined two metrics as proxy for diversity of expertise: 1) the maximum value of cosine distance between the authors, and 2) the number of connected components obtained when linking authors according to a similarity threshold.

---
[1] AIDA KG: http://aida.kmi.open.ac.uk/

The results show that both diversity metrics are significantly associated with the number of citations at five years. In other words, research papers authored by a research team with a wide set of skills and expertise tend to have a higher impact than the ones authored by more homogeneous teams.

The remainder of the paper is organised as follows. Section 2 provides an overview of the state of the art, while Section 3 outlines the materials and methodologies employed in the study. The findings are presented in Section 4. Section 5 concludes the paper by summarizing the main insights and outlining future directions.

## 2. Literature Review

In the literature, we can find a plethora of studies that analysed research team diversity across several dimensions: nationality (Smith et al., 2014), ethnicity (AlShebli et al., 2018; Freeman & Huang, 2015), institutions (Jones et al., 2008), gender (Nielsen et al., 2017), academic age (Jones & Weinberg, 2011), disciplinary backgrounds (Uzzi et al., 2013), and team size (Wu et al., 2019). The literature consensus is that a higher diversity often leads to an increase in productivity or impact. For instance, Smith et al. (2014) showed that promoting international collaboration has important benefits for scientific visibility, quality, and impact. Likewise, research on cross-institution teams in the field of engineering, social science, and others highlighted that multi-university collaborations with top-tier universities produce high-impact papers (Jones et al., 2008). Wu et al. (2019) analysed the size of research teams and found that small teams tend to build on less popular and potentially disruptive ideas, but also experience a citation delay, whereas larger teams work on more popular ideas and gather citations rapidly (Wu et al., 2019). AlShebli et al. (2018) studied the effect of ethnicity, gender, academic age, and affiliations on research impact. Their analysis shows that, even if all these factors play a role, ethnicity is the most prominent one, associated with an impact gain of 10.63%.

In this manuscript, we focus instead on the diversity of expertise, which has been notoriously hard to study since scholarly datasets lack a high-quality representation of researchers' expertise.

## 3. Materials and Methods

In this section, we describe the data source and the methodologies used to assess the diversity of expertise within a paper.

### 3.1. Data source - AIDA Knowledge Graph

The Academia/Industry DynAmics Knowledge Graph (AIDA KG) is a knowledge graph that describes 21 million publications and 8 million patents, in the field of Computer Science (Angioni et al., 2021). All documents are classified with respect to highly granular research topics, drawn from the Computer Science Ontology (Salatino et al., 2018, 2019). AIDA KG further characterizes 4.5 million publications and 5 million patents according to the type of the author's affiliations (academia, industry, or collaborative) and 66 industrial sectors (e.g., automotive, financial, energy, electronics) organized in a two-level taxonomy. It was generated by an automatic pipeline that integrates data from Microsoft Academic Graph, Dimensions,

DBpedia, the Computer Science Ontology, and the Global Research Identifier Database (GRID). It is publicly available under CC BY 4.0 and can be downloaded as a dump[2] or queried via a triplestore[3].

*3.2. Data Selection*

To analyse whether expertise diversity is related to the number of citations, we selected 114,203 research publications fulfilling four constraints: i) they were published between 2010 and 2015, ii) they reached at least 2 citations in the following five years, iii) they were authored by at least two authors, and iv) each author had at least one publication in the five years prior the paper under analysis. We set the first constraint to compare papers in the same time period. The second condition excludes patents and other technical documents that sometimes get included in the dataset, but do not typically receive citations. The third condition is a minimum requirement to analyse the characteristics of a research team. The last condition is required to compute metrics that consider the recent expertise of the authors. In practice, we first randomly selected 150K papers from AIDA KG in the period 2010-15 (first constraint) and then removed the ones that did not meet the remaining constraints.

We assessed the impact of a paper according to the number of citations received 5 years after its publication. For instance, the impact of a paper published in 2013 would be based on all the citations gathered by 2018. We then split the papers in 10 buckets of papers according to their number of citations after 5 years. Table 1 reports these groups alongside their frequency and the citation median.

Table 1. Groups of papers according to the citation ranges.

| Bucket Identifier | Citation ranges (c) | Citation Median | Num. of Papers |
|---|---|---|---|
| A | $2 \leq c < 5$ | 3 | 37,232 |
| B | $5 \leq c < 10$ | 6 | 27,696 |
| C | $10 \leq c < 15$ | 12 | 12,606 |
| D | $15 \leq c < 20$ | 17 | 7,180 |
| E | $20 \leq c < 30$ | 24 | 7,355 |
| F | $30 \leq c < 40$ | 34 | 3,717 |
| G | $40 \leq c < 50$ | 44 | 2,181 |
| H | $50 \leq c < 100$ | 64 | 3,691 |
| I | $100 \leq c < 150$ | 118 | 6,245 |
| J | $c \geq 150$ | 226 | 6,292 |

*3.3. Assessing Author Expertise*

As a following step, we identified 363,381 authors from the 114K papers and determined their expertise. Specifically, for each author, we selected their research publications in the 5 years prior to the publication of the paper under analysis. Next, we computed the distribution of topics in these articles, i.e., we counted the times a given topic appeared in the relevant papers. We normalised this distribution over the total number of papers and subtracted the normalised topic distribution of the whole Computer Science domain. This was done to identify the topics that are more relevant to the specific author, as suggested in (Angioni et al., 2022). For instance, the final weight of the topic *Machine Learning* for an author will be 40% if that topic appears in 70% of their articles and in 30% of Computer Science papers.

---

[2] AIDA KG dump: http://aida.kmi.open.ac.uk/downloads
[3] Triplestore of AIDA KG: https://aida.kmi.open.ac.uk/sparql

Finally, we ranked the topics based on this score and selected the top 10. We ran different experiments by testing other values between 5 and 20, but the overall results were very similar.

*3.4. Assessing Expertise Diversity in a Team of Authors*

To assess the diversity of expertise, we computed two statistical metrics on each paper. The first is the maximum value of the cosine distances computed on each couple of authors in the research team. The second metric counts the number of sub-teams having different expertise.

Specifically, for a given research paper, we computed the cosine distance between each couple of authors based on their top-10 topics, generating a distribution of $(N \times (N-1))/2$ values. The cosine distance is computed as the complement to one of the *cosine similarity*, and moves from 0 to 1. The higher the cosine distance the more diverse the set of topics between the two authors.

In this context, the average of the cosine distance is a bad indicator, since can produce very different results for papers that a researcher would consider very similar in terms of diversity. As an example, a research team consisting of an author in *Human Computer Interaction* and a second author in *Machine Learning* may obtain a fairly high value. However, a team composed of three authors in *Human-Computer Interaction* and three others in *Machine Learning*, would produce a much lower value. Therefore, we instead used as the first diversity metric the maximum value of the distribution, which does not suffer from this issue.

In order to produce a more granular metric that would reflect the different components in the team (two in the previous example), we clustered authors according to their expertise and counted the resulting number of subgroups. The higher the number of subgroups the more diverse the pool of researchers. Specifically, for each paper, we created an authorship graph $G = (V, E)$, where V is the set of authors and E is the list of edges. We generated an edge between a pair of authors when their cosine distance is below 0.3, i.e., they have a similarity higher than 0.7, which typically indicates a high degree of similarity between two vectors. Next, we extract the number of *connected components*, which are the groups of authors with similar expertise. Finally, based on the number of extracted components, we characterised the paper's diversity of expertise as: i) *low,* with 1 or 2 components, ii) *moderate,* with 3 or 4 components, iii) *high,* with 5 or 6 components, or iv) *very high,* from 7 components upward. Figure 1 shows an example of an author network with 7 authors arranged in 3 subgroups.

Figure 1. Example of network of authors with 3 components. In this case, the paper will be characterised as having moderate diversity.

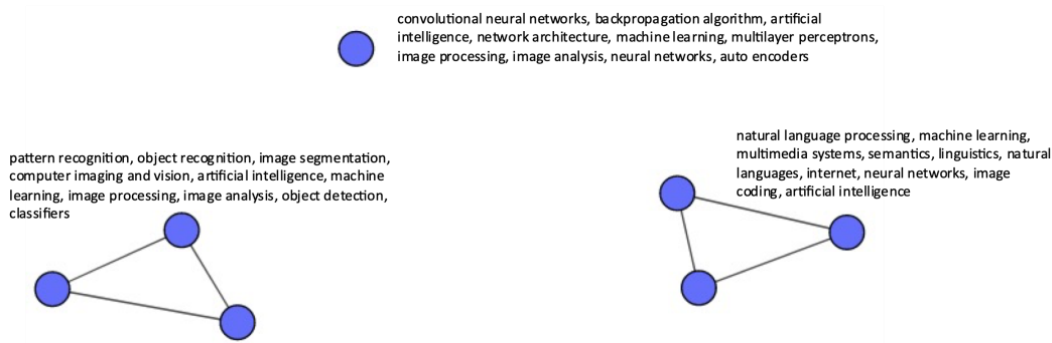

*3.5. Investigating the relationship between diversity of expertise and citations*

In order to assess if the expertise diversity of the authors of a paper is significantly associated with the number of citations in the following five years, we studied the distribution of the two previously described metrics across the 10 buckets. The difference between variables was studied with the chi-square test. The correlation between distributions of continuous variables was expressed by Pearson's linear correlation coefficient r, and relative p-value. Statistical significance corresponded in both cases to p<0.05.

## 4. Results
In this section, we discuss our results and report relevant statistical tests.

### 4.1. Max cosine distance between authors
Figure 2 reports the frequency of papers for a certain maximum cosine distance over the full dataset. The most notable characteristic is the peak at 1.

The specific distributions of the 10 buckets are similar in the range 0.1-0.9, but exhibit remarkable differences among them in the frequencies of 0 and 1. Therefore, we focused our analysis on these two cases. A score of 0 means that all the authors of a paper have exactly the same expertise (e.g., they all work on the very same branch of *Human-Computer Interaction*), while a score of 1 means that at least one author is working on completely different areas. The ratio #1/#0 can thus be used as a good indication of diversity. A higher ratio of #1/#0 will point to a higher expertise diversity.

Figure 2. The distribution of the maximum values of cosine distance.

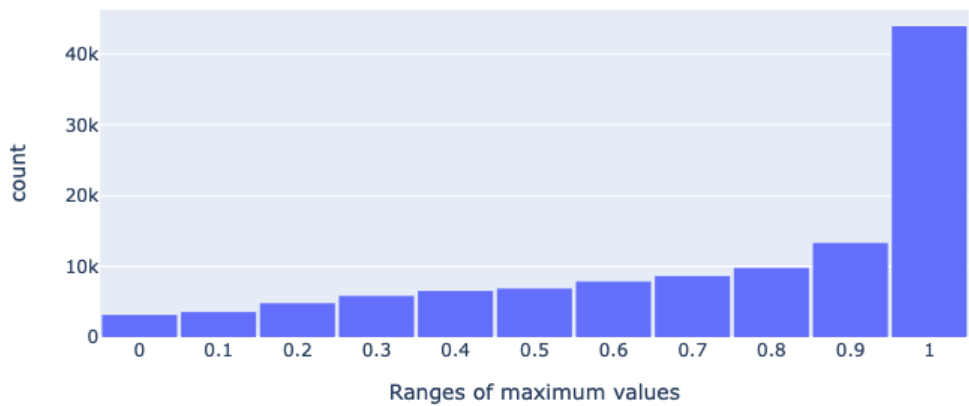

Table 2 reports on the number of 0 and 1 across the buckets as well as the ratio between these two values. Figure 3 further highlights this phenomenon by showing the #1/#0 ratio against the citation median of each bucket. The Pearson correlation coefficient between the distributions of these two variables is 0.955 (p<0.0001), which represents a strong direct linear correlation. This seems to confirm the hypothesis that a higher expertise diversity leads to a higher number of citations, i.e., to a higher impact.

Table 2. Frequency of research papers with zeros and ones according to the ranges of citations.

| Identifier | # of 0s | # of 1s | #1/#0 |
|---|---|---|---|
| A | 1,195 | 14,401 | 12.05 |
| B | 578 | 10,726 | 18.56 |
| C | 189 | 4,809 | 25.44 |
| D | 96 | 2,689 | 28.01 |
| E | 71 | 2,787 | 39.25 |
| F | 32 | 1,415 | 44.22 |
| G | 23 | 820 | 35.65 |
| H | 28 | 1,398 | 49.93 |
| I | 33 | 2,406 | 72.91 |
| J | 25 | 2,351 | 94.04 |

Figure 3. The #1/#0 ratio against the number of citations on a logarithmic scale.

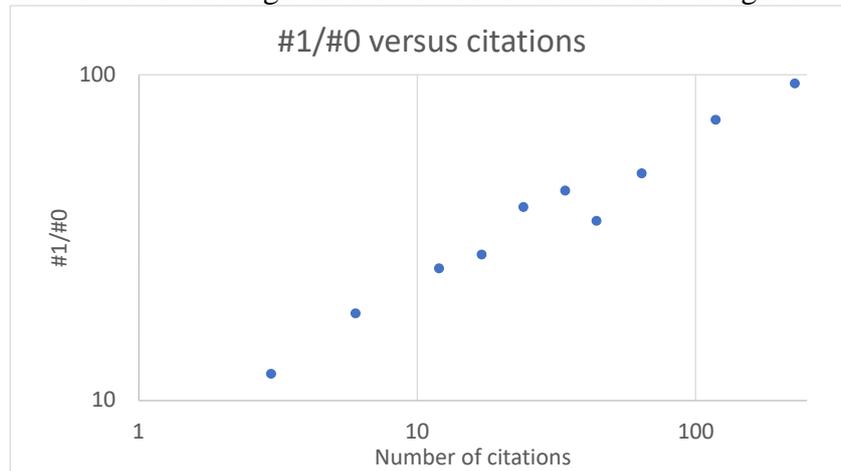

*4.2. Number of components in the author graph*

Table 3 reports the distribution of articles with low, moderate, high, and very high diversity as defined in Section 3.4. The percentage of papers with low diversity is inversely correlated to the median number of citations (r= -0.80, p=0.03), whereas the number of papers with high diversity is directly correlated (r=0.97, p<0.0001). Hence, the ratio between the number of papers with high or very high diversity and the ones with low diversity (see last column of Table 3) shows a significantly high direct correlation (Pearson's r=0.90) with the median number of citations. For instance, the set of papers with less than five citations (A) includes only 3.0% of the articles with high or very high diversity, while the set of high-impact papers with more than 150 citations (J) includes 9.7% of them.

The chi-square test applied the distribution of the four diversity categories (low, moderate, high, and very high) between adjacent buckets $i$ and $(i+1)$ found a significant difference between A and B and B and C (p<0.0001), decreasing for C vs D (p<0.04), and becoming not significant for the following pairs. A is also significantly different from B-J (p<0.0001) and A-B is significantly different from C-J (p<0.0001).

Figure 4 further showcases this dynamic by reporting the difference in the ratio of the diversity categories between the B-J buckets and A. For instance, the difference between the ratio of high-diversity papers in J (7.99%) and A (2.78%) is 5.21%.

In conclusion, the results obtained by using the four categories of diversity align with the ones based on the maximum cosine distance. In both cases, the expertise diversity metric is significantly associated with the number of citations.

Table 3. Percentages of papers with low, moderate, high, and very high diversity.

|   | low | moderate | high | very | Total | $\frac{very\ high + high}{low}$ |
|---|---|---|---|---|---|---|
| A | 64.84% | 32.15% | 2.79% | 0.23% | 37,232 | 0.05 |
| B | 61.69% | 34.71% | 3.25% | 0.35% | 27,700 | 0.06 |
| C | 60.06% | 35.40% | 4.14% | 0.40% | 12,606 | 0.08 |
| D | 58.23% | 36.56% | 4.75% | 0.46% | 7,180 | 0.09 |
| E | 57.92% | 36.56% | 4.88% | 0.64% | 7,355 | 0.10 |
| F | 56.60% | 37.18% | 5.62% | 0.59% | 3,717 | 0.11 |
| G | 56.44% | 37.37% | 5.64% | 0.55% | 2,181 | 0.11 |
| H | 54.67% | 37.83% | 6.52% | 0.97% | 3,695 | 0.14 |
| I | 52.49% | 39.12% | 7.21% | 1.18% | 6,245 | 0.16 |
| J | 51.16% | 39.16% | 7.99% | 1.68% | 6,292 | 0.19 |
| ALL | 60.51% | 34.91% | 4.10% | 0.49% | 114,203 | 0.08 |

Figure 4. Difference in the ratio of the diversity categories between the B-J buckets and A.

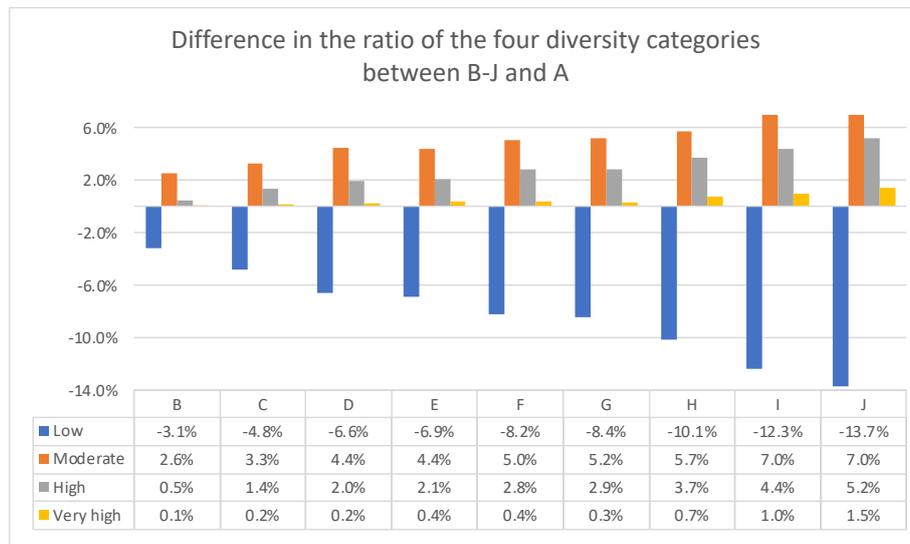

## 5. Conclusions and Future Work

In this study, we investigated whether the diversity of expertise in a research team can influence their scientific impact, measured as the number of citations received by the resulting papers in the upcoming 5 years. To this purpose, we represented the researcher's expertise according to a set of topics drawn from a fine-grained taxonomy of 14K research topics in the field of Computer Science. We then assessed the expertise diversity of a team by means of two metrics: i) the maximum cosine distance between the authors, and ii) the number of connected components obtained when linking authors according to a similarity threshold. Our experiments on a set of 114K papers show that both diversity metrics are significantly associated with the number of citations at five years. This suggests that, at least in Computer Science, diversity of expertise is key to scientific impact.

In future work, we plan to investigate additional metrics to measure the diversity of expertise in science. We plan also to expand our analysis to other fields of science, such as Engineering, Materials Science, and Medicine.

**Open science practices**
The analyses presented in this manuscript fully adhere to the open science practices. The main data source (the Academia/Industry DynAmics Knowledge Graph) is publicly available under a Creative Commons Attribution 4.0 International License (CC BY 4.0), and it can be downloaded from: http://aida.kmi.open.ac.uk/downloads. In addition, the sample data, code, and results are made available through Zenodo: https://doi.org/10.5281/zenodo.7846548.

**Acknowledgments**
We would like to thank Springer Nature for funding this research.

**Author contributions**
Angelo Salatino: Conceptualization, Data Curation, Methodology, Software, Writing - Review & Editing.
Simone Angioni: Conceptualization, Data Curation, Software, Resources.
Francesco Osborne: Project Administration, Supervision, Conceptualization, Methodology, Software, Writing - Review & Editing.
Diego Reforgiato Recupero: Project Administration, Supervision, Methodology, Writing - Review & Editing.
Enrico Motta: Project Administration, Supervision, Methodology, Writing - Review & Editing.

**Competing interests**
The authors declare that they have no competing interests.